\documentclass[aps,prl,twocolumn,amsmath,10pt,superscriptaddress,floatfix,nofootinbib]{revtex4-1}

\usepackage{epsfig,amssymb,amsfonts,amsmath,mathtools,bm,color,xcolor,graphicx,braket,adjustbox,esint,upgreek}
\usepackage{multirow}
\usepackage[utf8]{inputenc}

\newcommand{\be}{\begin{equation}}
\newcommand{\ee}{\end{equation}}
\newcommand{\bea}{\begin{eqnarray}}
\newcommand{\eea}{\end{eqnarray}}
\newcommand{\beas}{\begin{eqnarray*}}
\newcommand{\eeas}{\end{eqnarray*}}

\def\vec#1{\boldsymbol{#1}}

\newcommand{\x}{X_0(2900)}

\newcommand{\scnu}{\affiliation{{Guangdong Provincial Key Laboratory of Nuclear Science,}\\ Institute of Quantum Matter, South China Normal University, Guangzhou 510006, China}}
\newcommand{\tpcsf}{\affiliation{Institute of High Energy Physics, Chinese Academy of Sciences, Beijing 100049, China}}

\newcommand{\snsc}{\affiliation{Guangdong-Hong Kong Joint Laboratory of Quantum Matter, \\Southern Nuclear Science Computing Center, South China Normal University, Guangzhou 510006, China}}

\graphicspath{{FiguresV2/}}

\begin{document}

\title{The $X_0(2900)$ and its heavy quark spin partners in molecular picture}

\author{Mei-Wei Hu}
\scnu

\author{Xue-Yi Lao}
\scnu

\author{Pan Ling}
\scnu

\author{Qian Wang}\email{qianwang@m.scnu.edu.cn}
\scnu\snsc\tpcsf

\begin{abstract}
The $X_0(2900)$ observed by the LHCb Collaboration recently
in the $D^-K^+$ invariant mass of the $B^+\to D^+D^-K^+$ process is
the first exotic candidate with four different flavors, which opens a new era 
for the hadron community. Under the assumption that the $X_0(2900)$
 is a $I(J^P)=0(0^+)$ $\bar{D}^*K^*$ hadronic molecule, we extract
 the whole heavy-quark symmetry multiplet formed by the $\left(\bar{D},\bar{D}^*\right)$
 doublet and the $K^*$ meson.  
 For the bound state case, 
 there would be two additional $I(J^P)=0(1^+)$ hadronic 
  molecules associated with the $\bar{D}K^*$ and $\bar{D}^*K^*$ channels
   as well as one additional $I(J^P)=0(2^+)$ $\bar{D}^*K^*$ molecule.
    In the light quark limit, they are $36.66~\mathrm{MeV}$
and $34.22~\mathrm{MeV}$ below the $\bar{D}K^*$ and $\bar{D}^*K^*$ thresholds,
respectively, which are unambiguously fixed by the mass position of the $X_0(2900)$. 
 For the virtual state case, 
 there would be one additional $I(J^P)=0(1^+)$ hadronic 
  molecule strongly coupling to the $\bar{D}K^*$ channel
   and one additional $I(J^P)=0(2^+)$ $\bar{D}^*K^*$ molecule.
   Searching for these heavy quark spin partners will help shed light on the nature of the $X_0(2900)$.
\end{abstract}

\maketitle

{\it Introduction.}~~ 
The conventional quark model~\cite{GellMann:1964nj,Zweig:1981pd}, which inherits
part of the properties of Quantum Chromo-Dynamics (QCD), has made a great success
to understand hadrons before 2003.  Quark model tells us that hadrons can be classified as either mesons made of $q\bar{q}$
or baryons made of three quarks. However, QCD tells us that any color neutral configuration (especaily exotics) could exist
upon the two configurations mentioned above.  That leaves us two questions:
where to find these exotic candidates and how to understand the underlying mechanism.
 The observation of the first exotic candidate $X(3872)$~\cite{Choi:2003ue} in 2003 and the succeed tremendous experimental measurements ~\cite{Olsen:2017bmm,Yuan:2019zfo} 
 partly answer the first question. Among these, the observation of the first pentaquarks~\cite{Aaij:2015tga,Aaij:2019vzc},
 the first fully heavy four quark states~~\cite{LHCb-X6900}  and the first exotic candidates with four different flavors
 i.e. the $X_0(2900)$~\cite{Aaij:2020hon,Aaij:2020ypa} reported by the LHCb Collaboration recently,
  set milestones from experimental side. Different prescriptions 
  from theoretical side are put forward for understanding the nature of these exotic candidates
  ~\cite{Guo:2017jvc,Chen:2016qju,Esposito:2016noz,Ali:2017jda,Liu:2019zoy,Brambilla:2019esw,Bondar:2016hva,Lebed:2016hpi}. 
  Among them, hadronic molecule~\cite{Guo:2017jvc},  as an analogy of deuteron formed by a proton and a neutron, is proposed
  due to the fact that they are with a few MeV below the nearby $S$-wave threshold.

However, one have to confront one problem that
 different configurations with the same quantum number can mix with each other
 and cannot be well isolated. For instance,  although the $X(3872)$
 is proposed as a hadronic molecule at the very beginning~\cite{Tornqvist:2004qy}
 due to its closeness to the $D\bar{D}^*+c.c.$ threshold,
 it still could be or mix with the normal charmonium $\chi_{c1}(2P)$~\cite{Meng:2007cx,Kalashnikova:2005ui,Zhang:2009bv,Danilkin:2010cc,Li:2009zu,Li:2009ad,Coito:2010if,Coito:2012vf}.
 Another typical example is the $D^*_{s0}(2317)$ and the $D_{s1}(2460)$  which are
 about $160~\mathrm{MeV}$ and $70~\mathrm{MeV}$ below the $J^P=0^+$ and $J^P=1^+$ $c\bar{s}$ charmed-strange mesons of
Godfrey-Isgur quark model~\cite{Godfrey:1985xj}. Meanwhile, they are about $45~\mathrm{MeV}$ below the $DK$ and $D^*K$,
thresholds, respectively, which can be explained naturally if the
 systems are bound states of the $DK$ and $D^*K$ meson pairs
 ~\cite{Barnes:2003dj,vanBeveren:2003kd,vanBeveren:2003jv,Kolomeitsev:2003ac,Guo:2006fu,Zhang:2006ix,Guo:2006rp,Guo:2009id}.
 However, because the light quark and anti-quark in the isosinglet $D^{(*)}K$ system
are of the same flavor, 
 despite of those comprehensive studies, one still cannot avoid the possibility of the mixture with the normal $c\bar{s}$ configurations
 ~\cite{vanBeveren:2003kd,vanBeveren:2003jv,Coito:2011qn,Hwang:2004cd,Simonov:2004ar,Lee:2004gt,Zhou:2011sp}.
 Fortunately, the LHCb Collaboration reported a $J^P=0^+$~\cite{Aaij:2020hon,Aaij:2020ypa}
 narrow state $\x$ with mass $2866\pm 7~\mathrm{MeV}$ and width $\Gamma_0=57\pm 13~\mathrm{MeV}$
 as well as another broader $J^P=1^-$ state with mass $2904\pm 7~\mathrm{MeV}$ and width $\Gamma_1=110\pm 12~\mathrm{MeV}$
 in the $\bar{D}K$ invariant mass distribution. They are the first exotic states with four different flavors,
which brings us a potential ultimate solution for the problem from different aspects.

In this letter, we solve the Lippmann-Schwinger Equation (LSE) with leading order contact potentials of 
the $\bar{D}^{(*)}K^{(*)}$ system, in the heavy quark limit, to extract the mass position of the spin partners of the $X_0(2900)$.
That the $\x$ exists as a $I(J^P)=0(0^+)$ $\bar{D}^*K^*$ hadronic molecule is an input in our framework.
With that assumption, we predict the masses of its heavy quark spin partners.
Searching for those spin partners could help to understand the nature of the $X_0(2900)$.

\medskip
{\it Framework.}~~ 
The heavy quark spin structure~\cite{Bondar:2011ev} could reexpress of the hadron basis by
the heavy-light basis. One could find an example for the 
  $Z^{(\prime)}_c$ and $Z_b^{(\prime)}$ case with two heavy quarks in Refs.~\cite{Hanhart:2015cua,Guo:2016bjq,Wang:2018jlv,Baru:2019xnh,Voloshin:2011qa,Baru:2017gwo}.
   Along the same line, the $S$-wave $\bar{D}^{(*)}K^{(*)}$ system with only one heavy quark can be
written in terms of the heavy degree of freedom $\frac{1}{2}$ and light degree of freedom $s_l$
as the following~\cite{Yasui:2013vca}
\begin{widetext}
\begin{equation}
|(\bar{c}_{j_{1}}q_{j_{2}})_{j_{12}}(\bar{s}q^{\prime})_{j_{3}}\rangle_{J}=\sum_{s_{l}}\left(-1\right)^{j_{2}+j_{3}+j_{12}+s_{l}}\left\{ \begin{array}{ccc}
j_{1} & j_{2} & j_{12}\\
J & j_{3} & s_{l}
\end{array}\right\}\hat{j_{12}}\hat{s_l} |\bar{c}_{j_{1}}(q_{j_{2}}(\bar{s}q^{\prime})_{j_{3}})_{s_{l}}\rangle
\label{eq:decomposition}
\end{equation}
\end{widetext}
with $\hat{j}=\sqrt{2j+1}$.
Here, $j_1=\frac{1}{2}$, $j_2=\frac{1}{2}$ and $j_{12}=0,1$ are spins of anti-charm quark $\bar{c}$, light quark $q$ and the sum of them in the $\bar{D}^{(*)}$ meson.
$j_3=0,1$ and $J=0,1,2$ are the spins of the $K^{(*)}$ meson and the total spin of the $\bar{D}^{(*)}K^{(*)}$ system. 
$s_l$ on the right hand side of Eq.~\eqref{eq:decomposition} is the light degree of freedom of the system,
which is the only relevant quantity for the dynamics in the heavy quark limit. Following Eq.~\eqref{eq:decomposition},
one can obtain the decompositions of the $\bar{D}^{(*)}K^{(*)}$ system as
\begin{eqnarray}
|\bar{D}K\rangle_{0^+}&=&|\frac{1}{2}\rangle\\
|\bar{D}^*K\rangle_{1^+}&=&|\frac{1}{2}\rangle\\
|\bar{D}K^*\rangle_{1^+}&=&\frac{1}{\sqrt{3}}|\frac{1}{2}\rangle^*+\sqrt{\frac{2}{3}}|\frac{3}{2}\rangle^*\\
|\bar{D}^*K^*\rangle_{0^+}&=&-|\frac{1}{2}\rangle^*\\
|\bar{D}^*K^*\rangle_{1^+}&=&\sqrt{\frac{2}{3}}|\frac{1}{2}\rangle^*-\frac{1}{\sqrt{3}}|\frac{3}{2}\rangle^*\\
|\bar{D}^*K^*\rangle_{2^+}&=&|\frac{3}{2}\rangle^*.
\end{eqnarray}
Here the heavy degree of freedom is suppressed due to the same value, leaving only the light degrees of 
freedom $s_l$ in $|\dots\rangle$. Although the $K$ and $K^*$ have the same quark content,
the light degrees of freedoms in the first two equations and those in
the last four equations can be distinguishable due to the large scale separation of the $K$ and $K^*$ masses.
Analogous to those in Refs.~\cite{Guo:2013sya}, by defining the contact potential
\begin{eqnarray}
C_{2l}^{(*)}\equiv^{(*)}\langle l |\hat{H}_\mathrm{HQS}|l\rangle^{(*)},
\end{eqnarray}
the potentials of the $\bar{D}^{(*)}K$ and and $\bar{D}^{(*)}K^*$ systems are
\begin{eqnarray}
V_{0^+}&	=&C_{1}\label{eq:potential0}\\
V_{1^+}&	=&C_{1}\label{eq:potential1}\
\end{eqnarray}
and
\begin{eqnarray}
V^*_{0^+}&	=&C^*_{1}\\
V^*_{1^+}&	=&\left(\begin{array}{cc}
\frac{1}{3}C^*_{1}+\frac{2}{3}C^*_{3} & \frac{\sqrt{2}}{3}\left(C^*_{1}-C^*_{3}\right)\\
\frac{\sqrt{2}}{3}\left(C^*_{1}-C^*_{3}\right) & \frac{2}{3}C^*_{1}+\frac{1}{3}C^*_{3} 
\end{array}\right)\\
V^*_{2^+}&	=&C^*_{3},
\end{eqnarray}
respectively. 
$V_{J^+}$ and $V^*_{J^+}$ are for the potentials of the $\bar{D}^{(*)}K$ and $\bar{D}^{(*)}K^*$ systems, respectively.
The subindex $J^+$ presents the total spin and parity of the corresponding system.
The transition between $|l\rangle$ and $|l\rangle^*$ is the higher order contribution,
which is set to zero in this work~\footnote{The observation of the $X_0(2900)$ in the $\bar{D}K$ channel is 
due to this higher order contribution, i.e. the mixing between the $|\frac{1}{2}\rangle$ and the $|\frac{1}{2}\rangle^*$ compoments.}.
The above decomposition and the corresponding potentials also work for the $D^{(*)}K^{(*)}$ systems, but with different values of $C_{2l}^{(*)}$.
\begin{figure*}[!htb]
 \centering
  \includegraphics[width=0.45\textwidth]{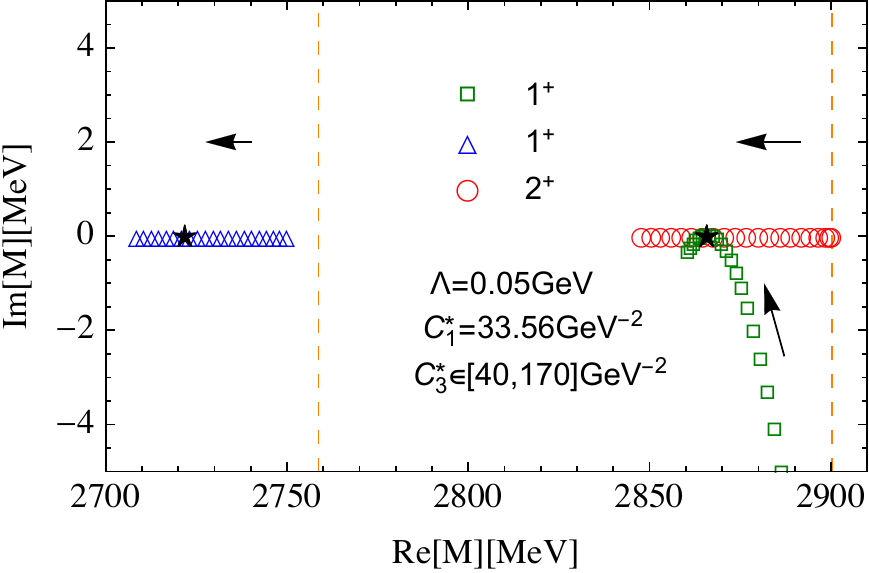}  \includegraphics[width=0.45\textwidth]{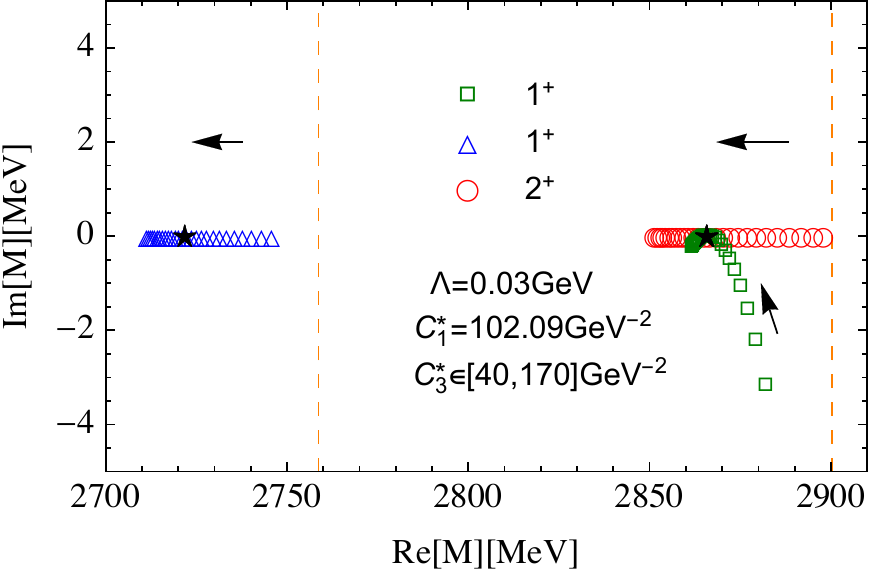}
  \caption{The $X_0(2900)$ is assumed to be a $I(J^P)=0(0^+)$
  $\bar{D}^*K^*$ bound state. The values $C_1^*=33.56~\mathrm{GeV}^{-2}$ 
  and $C_1^*=102.09~\mathrm{GeV}^{-2}$ are obtained for $\Lambda=0.05~\mathrm{GeV}$
  and $\Lambda=0.03~\mathrm{GeV}$, respectively.
   The $2^+$ (red circles) and the lower $1^+$ (blue triangles) behave as bound states.
   The higher $1^+$ (green squares) is a resonance between the $\bar{D}K^*$ and $\bar{D}^*K^*$ thresholds.
    The pole trajectories of these three states with the increasing $C_3^*$
    are shown in the figures.
  The black arrows indicate the direction of the increasing $C_3^*$.
  The black stars are the mass positions in the light quark spin symmetry.
        }
  \label{fig1}
\end{figure*}

With the above potentials, one can solve LSE
\begin{equation}
T=V+VGT
\end{equation}
with $V$ the potentials for specific channels of a given quantum number. 
Here two-body non-relativistic propagator is
\begin{eqnarray*}
&&G_\Lambda(M,m_1,m_2)=\int\frac{\mathrm{d}^3q}{(2\pi)^3}\frac{1}{M-m_1-m_2-\vec{q}^2/(2\mu)}\\
&&=\Lambda+i\frac{m_1 m_2}{2\pi(m_1+m_2)}\sqrt{2\mu (M-m_1-m_2)}
\end{eqnarray*}
with power divergence subtraction~\cite{Kaplan:1998tg} to regularize the ultraviolet (UV) divergence.
The value of $\Lambda$ should be smaller enough to preserve heavy quark symmetry,
 leaving the physics insensitive to the details of short-distance dynamics~\cite{Guo:2013sya}.
Here $m_1$, $m_2$ and $\mu$ are the masses of the intermediated two particles and their reduced mass, respectively.
 $M$ is the  total energy of the system. The expression of the second Riemann sheet $G^{\mathrm{II}}_\Lambda(M,m_1,m_2)$ 
 can be obtained by changing the sign of the second term of $G_\Lambda(M,m_1,m_2)$.

\medskip
{\it Results and Discussions.}~~
Before the numerical results, we estimate the values of the contact potential $C_1$.
The leading contact terms between heavy-light mesons and Goldstone bosons can be obtained by the following Lagrangian
~\cite{Guo:2008gp,Du:2017ttu,Burdman:1992gh,Wise:1992hn,Yan:1992gz,Yao:2015qia}
 \begin{eqnarray}
 \mathcal{L}_{D\phi}^{(1)}=\mathcal{D}_{\mu}D\mathcal{D}^{\mu}D^{\dagger}-M_{0}^{2}DD^{\dagger},
 \label{eq:lag}
  \end{eqnarray}
  where 
  \begin{eqnarray}
\mathcal{D}_{\mu}H=H\left(\overset\leftarrow{\partial}_{\mu}+\Gamma_{\mu}^{\dagger}\right),\quad\mathcal{D}_{\mu}H^{\dagger}=\left(\partial_{\mu}+\Gamma_{\mu}\right)H^{\dagger},
  \end{eqnarray}
  with 
   \begin{eqnarray}
H\in\left\{ D,D^{*}\right\} ,\quad D^{(*)}=\left(D^{(*)0},D^{(*)+},D_{s}^{(*)+}\right)
  \end{eqnarray}
  and chiral connection
 \begin{eqnarray}
\Gamma_{\mu}=\left(u^{\dagger}\partial_{\mu}u+u\partial_{\mu}u^{\dagger}\right)/2.
  \end{eqnarray}
  Here the chiral building blocks are 
 \begin{eqnarray}\nonumber
u_{\mu}=i\left[u^{\dagger}\partial_{\mu}u-u\partial_{\mu}u^{\dagger}\right],\quad U=u^{2},\quad\chi^{\pm}=u^{\dagger}\chi u^{\dagger}\pm u\chi^{\dagger}u.
  \end{eqnarray}
  Here $U=\exp\left(i\sqrt{2}\phi/f_0\right)$ with $\phi$ the Goldstone boson octet.
  To the leading order, $f_0$ is the pion decay constant. The isospin singlet $J^P=0^+$ $DK$ and $\bar{D}K$
  systems are defined as
  \begin{eqnarray}
D_{s0}^{*+}(2317)\equiv\frac{1}{\sqrt{2}}\left(D^{0}K^{+}-D^{+}K^{0}\right),
\end{eqnarray} 
which is associated with the $D_{s0}^{*+}(2317)$
and 
  \begin{eqnarray}
 X^\prime_{0}\equiv\frac{1}{\sqrt{2}}\left(D^{0}\bar{K}^{0}+D^{+}K^{-}\right).
\end{eqnarray} 
The definitions of the isospin singlet $J^P=1^+$ $D^*K$ and $\bar{D}^*K$ systems
are analogous. 
From Eq.~\eqref{eq:lag}, we obtain
\begin{eqnarray}
V_{D_{s0}^{*+}(2317)}&=&V_{D_{s1}^{+}(2460)}
\end{eqnarray}
which agrees with those obtained from the heavy-light decomposition, i.e. Eqs.~\eqref{eq:potential0},~\eqref{eq:potential1},
and
\begin{eqnarray}
V_{D_{s0}^{*+}(2317)}&=&2V_{X_0}.\label{eq:relation}
\end{eqnarray}
As the result, the value of $C_1$ for the $\bar{D}^{(*)}K$ system is half of that for the $D^{(*)}K$ system. 
We find that any parameter set, i.e. $(\Lambda, C_1)$, for the existence of the $D_{s0}^*(2317)$ and $D_{s1}(2460)$ as $DK$ and $D^*K$ molecular states
(both bound states and virtual states) does not indicate the existence of the analogous $\bar{D}K$ and $\bar{D}^*K$ molecules. 
Here and in what follows, we focus on the discussion of the formation of the 
$\bar{D}^{(*)}K^{(*)}$ molecule instead of their isospin breaking effect. As the result,
the isospin average masses 
\begin{eqnarray}
m_D&=&1.867~\mathrm{GeV},\quad m_{D^*}=2.009~\mathrm{GeV}\\
m_K&=&0.496~\mathrm{GeV},\quad m_{K^*}=0.892~\mathrm{GeV}
\end{eqnarray}
are implemented in this letter. 
\begin{figure*}[!htb]
 \centering
  \includegraphics[width=0.45\textwidth]{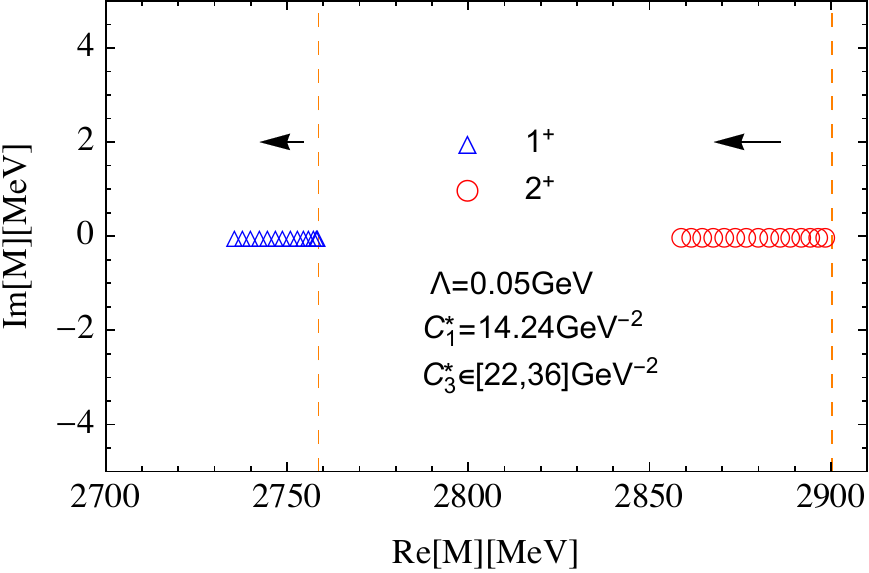}  \includegraphics[width=0.45\textwidth]{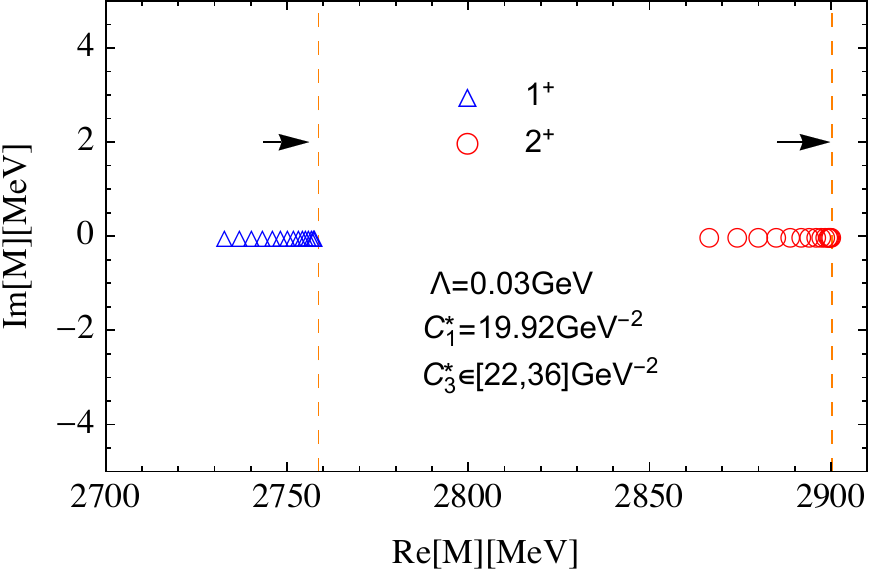}
  \caption{The $X_0(2900)$ is assumed to be a $I(J^P)=0(0^+)$
  $\bar{D}^*K^*$ virtual state. The values $C_1^*=14.24~\mathrm{GeV}^{-2}$ 
  and $C_1^*=19.92~\mathrm{GeV}^{-2}$ are obtained for $\Lambda=0.05~\mathrm{GeV}$
  and $\Lambda=0.03~\mathrm{GeV}$, respectively. 
   The $2^+$ (red circles) and the lower $1^+$ (blue triangles) behave as bound states for the former case.
   They behave as virtual states for the later case.
   The higher $1^+$ (green squares) is far away from the physical sheet and has marginal physical impact for both cases.
    The pole trajectories of the two states with the increasing $C_3^*$
    are shown in the figures.
  The black arrows indicate the direction of the increasing $C_3^*$.
        }
  \label{fig2}
\end{figure*}

For the $\bar{D}^{(*)}K^*$ interaction, the $X_0(2900)$ recently observed by the LHCb collaboration
is assumed to be a $I(J^P)=0(0^+)$ $\bar{D}^*K^*$ molecular state~\cite{Molina:2010tx}. We consider the two cases for the 
$X_0(2900)$
\begin{itemize}
\item a bound state with the $X_0(2900)$ mass $m_{X_0(2900)}$ satisfying 
\begin{equation}
1-C_1^* G_\Lambda(m_{X_0(2900)},m_{\bar{D}^*},m_{K^*})=0
\end{equation}
\item a virtual state with the $X_0(2900)$ mass $m_{X_0(2900)}$ satisfying 
\begin{equation}
1-C_1^* G^{\mathrm{II}}_\Lambda(m_{X_0(2900)},m_{\bar{D}^*},m_{K^*})=0
\end{equation}
\end{itemize}
We take $\Lambda=0.05~\mathrm{GeV}$ and $\Lambda=0.03~\mathrm{GeV}$
to illustrate the mass positions of its heavy quark spin partners and the corresponding properties.

For the bound state solution of the $X_0(2900)$, $C_1^*=33.56~\mathrm{GeV}^{-2}$ and $C_1^*=102.09~\mathrm{GeV}^{-2}$
correspond to $\Lambda=0.05~\mathrm{GeV}$ and $\Lambda=0.03~\mathrm{GeV}$, respectively.
Fig.~\ref{fig1} shows how the poles move with the variation of the two parameter sets.
The blue triangle and green square curves show the pole trajectory of
the bound state and resonance in the $1^+$ channel. One can see that, with $C^*_3$ variation between  
$40~\mathrm{GeV}^{-2}$ and $170~\mathrm{GeV}^{-2}$,
one bound state and one resonance emerge with tens of MeV
below the $\bar{D}K^*$ and $\bar{D}^*K^*$ thresholds, respectively. The bound state in the $2^+$
channel is more sensitive to the parameter $C_3^*$.
 If one would expect that light quark spin symmetry also works here as that for the two $Z_b$ states~\cite{Voloshin:2016cgm},
i.e. $C^*_3=C^*_1$, we find the
pole position of the above three states are
\begin{eqnarray}
m_{2^+}=2.866~\mathrm{GeV},\\
m_{1^+}=2.722~\mathrm{GeV},\quad m_{1^+}=2.866~\mathrm{GeV}
\end{eqnarray}
The vanishing imaginary part of the higher $1^+$ state is because of the degenerance of
the two $1^+$ states. 

For the virtual state solution of the $X_0(2900)$,
 $C_1^*=14.24~\mathrm{GeV}^{-2}$ and $C_1^*=19.92~\mathrm{GeV}^{-2}$
correspond to $\Lambda=0.05~\mathrm{GeV}$ and $\Lambda=0.03~\mathrm{GeV}$, respectively.
Fig.~\ref{fig2} shows how the poles move with $C^*_3$ variation between  
$22~\mathrm{GeV}^{-2}$ and $36~\mathrm{GeV}^{-2}$ for the two $\Lambda$ values.
For the former case, the blue triangles and red circles show the pole trajectories of the bound states
for the $1^+$ and $2^+$ channels, respectively. For the later case, both of them become virtual states. 
As the result, whether the higher $1^+$ $\bar{D}^*K^*$ molecule exists or not depends on the nature of the $X_0(2900)$,
i.e. either a bound state or a virtual state, which can be studied by the further detailed scan of its line shape. 
Thus, searching for these heavy quark spin partners would help to reveal
the nature of the $\x$.

\medskip
{\it Summary and Outlook.}~~
Under the assumption that the LHCb Collaboration recently
reported $\x$ is a $I(J^P)=0(0^+)$ $\bar{D}^*K^*$ hadronic molecule,
we extract the pole trajectories of its heavy quark spin partners with the variation of the parameter 
$C_3^*$. The parameter $C_1^*$ is fixed by the mass position of 
the $X_0(2900)$ (either a bound state or a virtual state). 
For the bound state case, in the light quark spin symmetry, we extract the mass positions of its heavy quark spin partners,
 i.e. $2.722~\mathrm{GeV}$ and $2.866~\mathrm{GeV}$ for $1^+$ state, 
and $2.866~\mathrm{GeV}$ for $2^+$ state. 
For the virtual state case, the higher $1^+$ state is far away from the physical region and will not have large impact on the physical observables.
Searching for those states would help to shed light on the nature of the $\x$.

During the update of this manuscript,
 several works
 \cite{He:2020jna,Liu:2020orv,Zhang:2020oze,Lu:2020qmp,Liu:2020nil,Chen:2020aos,Wang:2020xyc,Huang:2020ptc,Qin:2020zlg}
 appear to discuss the relevant topics.

\medskip

\begin{acknowledgements}
The discussions with Tim Burns, M.L.~Du, Li-Sheng Geng, Ming-Zhu Liu, Eulogio Oset, Jun-Jun Xie are appreciated. 
A special acknowledgement to C.~Hanhart for pointing out the relation between the $C_1$ and the potential of the $D_{s0}(2317)$ in hadronic
molecular picture to the leading order.
This work is partly supported by the National Natural Science Foundation of China (NSFC) and the Deutsche Forschungsgemeinschaft (DFG) through the funds provided to the Sino-German Collaborative Research Center “Symmetries and the Emergence of Structure in QCD” (NSFC Grant No.~11621131001 and DFG Grant No.~TRR110), Science and Technology Program of Guangzhou (No.~2019050001), NSFC Grant No.~12035007, Guangdong Provincial funding with Grant No.~2019QN01X172.
 MWH and XYL
are also supported by Entrepreneurship competition for College Students of SCNU.
\end{acknowledgements}

\end{document}